\documentclass[aps,prl,twocolumn,showpacs]{revtex4-1}

\usepackage{graphicx}
\usepackage{amsfonts}
\usepackage{amssymb}
\usepackage{amsmath}
\usepackage{mathrsfs}
\usepackage{braket}
\usepackage{color}

\begin{document}
\title{Counter-propagating edge modes and topological phases of a kicked quantum Hall system}
\author{Mahmoud Lababidi}
\affiliation{School of Physics, Astronomy, and Computational Sciences, George Mason University, Fairfax, VA 22030}
\author{Indubala I. Satija}
\affiliation{School of Physics, Astronomy, and Computational Sciences, George Mason University, Fairfax, VA 22030}
\author{Erhai Zhao}
\affiliation{School of Physics, Astronomy, and Computational Sciences, George Mason University, Fairfax, VA 22030}
\begin{abstract}
Periodically driven quantum Hall system in fixed magnetic field is found to exhibit a series of phases 
featuring anomalous edge modes with the ``wrong" chirality.
This leads to pairs of counter-propagating chiral edge modes at each edge, in sharp contrast
to stationary quantum Hall systems. We show the pair of Floquet edge modes 
are protected by the chiral (sublattice) symmetry, and that they are robust against static disorder.
The existence of distinctive phases with the same Chern 
and winding numbers but very different edge state spectra points to the important
role played by symmetry in classifying topological properties of driven systems. 
We further explore the evolution of the edge states with driving using a simplified model, 
and discuss their experimental signatures.
\end{abstract}
\maketitle
Cyclic time-evolutions of quantum systems are known to have interesting topological properties
 \cite{1984RSPSA.392.45B,PhysRevLett.58.1593}. 
Several groups recently showed  that
periodic driving can turn an ordinary band insulator 
(superconductor) into a Floquet topological insulator (superconductor) \cite{PhysRevB.79.081406,lindner_floquet_2011,
kitagawa_topological_2010,lindner_topological_2011,kitagawa_transport_2011,jiang_majorana_2011,
gu_floquet_2011,rudner_anomalous_2012,2013arXiv1303.2300T}. This provides a powerful way to engineer 
effective Hamiltonians that stroboscopically mimic stationary topological insulators \cite{lindner_floquet_2011,
kitagawa_topological_2010,Rechtsman:2013fk}. Moreover, a large class of topological phenomena in 
periodically driven many-body systems are unique and have no stationary counterparts.
An early example is Thouless's one-dimensional charge pump, where he showed 
that the charge transport is quantized and related to a topological invariant
 \cite{PhysRevB.27.6083}. Other topological invariants for the time evolution operator 
 in two and three dimensions have been constructed recently
\cite{PhysRevB.79.081406,kitagawa_topological_2010,rudner_anomalous_2012}.  
Yet a systematic classification of these invariants analogous to the periodic table of 
symmetry protected topological phases \cite{kitaev,PhysRevB.78.195125} is still to be achieved.

In this paper, we identify new topological phenomena 
in a lattice integer quantum Hall (QH) system under cyclic driving with period $T$. 
For fixed magnetic field, variations of the driving parameter induce topological 
phase transitions where the Chern numbers of the quasienergy
bands change. We find multiple phases of the driven QH system featuring 
counter-propagating chiral edge modes at the each edge, and show they
are robust against disorder. In particular, there
appear ``$\pi$-modes", pairs of edge modes with opposite chirality at quasienergy $\pi/T$.  
These anomalous edge modes differ from those found previously in other 
driven two-dimensional (2D) lattice models, where the edge modes at quasienergy $\pi/T$ 
all propagate in the same direction and subsequently their number can be inferred either
from the Chern number or the winding number
 \cite{kitagawa_topological_2010,rudner_anomalous_2012}.
Here, 
these known topological invariants can not predict the number of edge modes of each chirality, 
but only their difference. 
For example, we find two phases (phase A and D below) having the same 
set of Chern and winding numbers but very different edge state spectra. 
Our analysis suggests that symmetry of the time evolution operator 
has to be included to fully
characterize and understand the topological properties of driven systems.

Our work is motivated by 
recent experimental achievements
of artificial magnetic field for ultracold atoms \cite{Lin:2009fk,PhysRevLett.107.255301} and temporal modulation of 
optical lattices \cite{PhysRevLett.108.225304,PhysRevLett.109.145301}. 
We consider a model consisting of (spinless) fermionic atoms loaded onto a square optical lattice. 
Each site is labeled by vector ${\bf r}=n\hat{x}+m\hat{y}$, where $n$, $m$ are
integers, $\hat{x}$ ($\hat{y}$) is the unit vector in the $x$ ($y$) direction, and the lattice
spacing $a$ is set to be the length unit. The tight binding Hamiltonian has the form
\begin{equation}
H=-J_x\sum_{\bf r}|\mathbf{r}+\hat{x} \rangle\langle \mathbf{r}|
-J_y\sum_{\bf r}|\mathbf{r}+\hat{y} \rangle e^{i2\pi n\alpha} \langle \mathbf{r}|
+h.c. \label{qh}
\end{equation}
Here, $|\mathbf{r}\rangle$ is the Wannier state localized at site $\mathbf{r}$. $J_x$ ($J_y$)
is the nearest neighbor hopping along the $x$ ($y$) direction.
We assume a uniform synthetic magnetic field $B$ is applied in the $-z$ direction, and 
work in the Landau gauge, $A_x=0$, $A_y=-Bx$. The flux per plaquette, in units of
the flux quantum $\Phi_0$, is $\alpha=-Ba^2/\Phi_0$. Field $B$ gives
rise to the Peierls phase factor $e^{i2\pi n\alpha}$ in the hopping. For static $J_x$, $J_y$,
$H$ is the well known Hofstadter model \cite{hofstadter_energy_1976}.

We investigate a class of periodically driven quantum Hall systems 
described by $H$ above, but with $J_x$ and $J_y$ being periodic functions of time $t$. 
We will focusing on the following driving protocol
\begin{align}
J_x(t)=J_x,\;\; J_y(t)=0.\;\;\;  0<\mathrm{mod}(t,T)<\tau \nonumber \\
J_x(t)=0,\;\; J_y(t)=J_y.\;\;\;  \tau<\mathrm{mod}(t,T)<T \label{m1}
\end{align}
Namely, within one period $T$, the hopping along $x$ is turned on during the interval $(0,\tau)$, while
the hopping along $y$ is turned on during the interval $(\tau,T)$.
We then have two independent
driving parameters,
$\theta_x=J_x\tau/\hbar,\;\theta_y=J_y(T-\tau)/\hbar$.
While it is hard to achieve in solid state systems, 
temporal modulation of $J_x$ or $J_y$ is straightforward to implement for cold atoms in
optical lattices, e.g., by simply tuning the intensity of the laser.
In the limit $\tau\rightarrow T$ and $(T-\tau)J_y\rightarrow$const, the
driving protocol becomes
\begin{align}
J_x(t)=J_x,\;\; J_y(t)=J_yT\sum_j \delta(t-jT), \label{m2}
\end{align}
i.e., the $y$ hopping is only turned on when $t=jT$, with $j$ any integer. 
In this limit, $\theta_x=J_xT/\hbar$, $\theta_y=J_yT/\hbar$. 
We will simplify refer to systems described by \eqref{m1} or \eqref{m2} as 
kicked quantum Hall systems,
because \eqref{m2} resembles the well studied kicked rotors \cite{2013arXiv1303.2300T}.

The time evolution operator of the system, defined by $|\psi(t)\rangle=U(t)|\psi(0)\rangle$,
has the formal solution $U(t)={\cal T}\exp[-i\int^t_0H(t')dt']$,
where ${\cal T}$ denotes time-ordering and we set $\hbar=1$ throughout.
The discrete
translation symmetry $H(t)=H(t+T)$ leads to a convenient basis
$\{|\phi_\ell\rangle\}$, defined as the eigenmodes of Floquet operator $U(T)$,
\[
U(T)|\phi_\ell\rangle=e^{-i\omega_\ell T}|\phi_\ell\rangle.
\]
Here the quasienergy $\omega_\ell$, by definition, 
is equivalent to $\omega_\ell+2p\pi/T$ for any integer $p$ and lives
within the quasienergy Brillouin zone (QBZ), $\omega\in[-\pi/T,\pi/T)$.
%
For rational flux $\alpha=1/q$, 
$U$ is a $q\times q$ matrix in momentum space and there are 
$q$ quasienergy bands. For convenience, we label the lowest band within the QBZ with $\ell=1$,
and the subsequent bands at increasingly higher quasienergies with $\ell=2,3,...,q$.
Correspondingly, we call the gap below the $\ell$-th band the $\ell$-th gap.
For example, the gap around $\pm\pi/T$ is the first gap. The Chern number for the $\ell$-th
quasienergy band can be defined analogous to the stationary case \cite{thouless_quantized_1982}
\[
c_\ell=\frac{i}{2\pi }\int dk_x dk_y \left[
 \partial_{k_x} \phi^*_\ell(\mathbf{k}) \partial_{k_y} \phi_\ell(\mathbf{k}) - c.c.\right],
\]
where the integration is over the magnetic Brillouin zone, and
$\phi_\ell(\mathbf{k})$ is the $\ell$-th eigenwavefunction of $U(\mathbf{k},T)$.

\begin{figure}
\includegraphics[width=0.95\columnwidth]{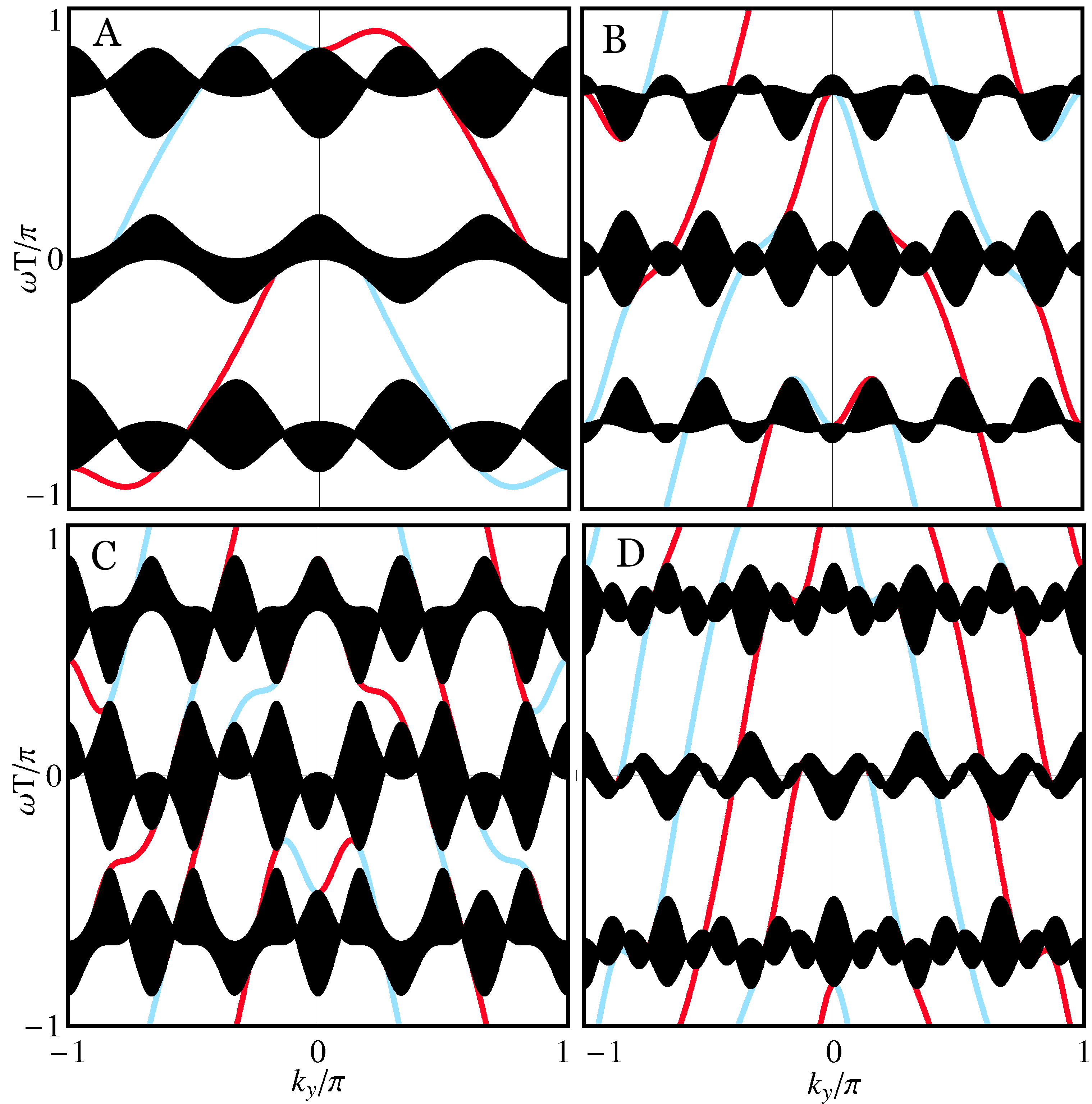}
\caption{(color online) Quasienergy spectra of a finite (in the $x$-direction) 
 slab of periodically driven quantum Hall system at flux $-1/3$ and fixed $\theta_x=\pi/3$. 
The four panels, $\theta_y=0.5\pi$, $\pi$, $1.2\pi$, and $1.5\pi$, correspond to
phase A, B, C, and D, respectively, shown in Fig. 2.
Edge states localized on the left (right)  edge are shown in blue (red).
}\label{spectrum}
\end{figure}

Figure 1 displays four representative quasienergy spectra of a finite slab of 
length $L$ in the $x$ direction under periodic driving \eqref{m1}. As in static QH systems,
we observe edge states forming within the quasienergy gaps.
Consider the left edge ($x=0$) and
let us denote the number of chiral edge modes propagating in the $\hat{y}$ ($-\hat{y}$) direction by 
$n_{\ell}^+$ ($n_{\ell}^-$). For driven 2D systems, the Chern numbers are generally insufficient to 
predict $(n_{\ell}^+,n_{\ell}^-)$. Instead, as shown by
Rudner et al \cite{rudner_anomalous_2012}, the net chirality of the edge modes inside the $\ell$-th
quasienergy gap, $w_\ell \equiv n_{\ell}^+-n_{\ell}^-$, is given by the following winding number 
\begin{align*}
w_\ell=\int \frac{dk_xdk_ydt}{24\pi^2} \epsilon^{\mu\nu\rho}\mathrm{Tr} \left[(u^{-1}\partial_\mu u) 
(u^{-1}\partial_\nu u)
 (u^{-1}\partial_\rho u)\right].
\end{align*}
Here $\mu,\nu,\rho=1,2,3$ corresponds to $k_x,k_y,t$ respectively, and $u(\mathbf{k},t)$
is a smooth extrapolation of $U(\mathbf{k},t)$ \cite{rudner_anomalous_2012}
\[
u(\mathbf{k},t)=U(\mathbf{k},2t)\theta(T/2-t)+e^{-i\mathscr{H}(\mathbf{k})2(T-\tau)}\theta(t-T/2),
\]
where $\mathscr{H}(\mathbf{k})=-(i/T)\log U(T)$ is the effective Hamiltonian with
the branch cut of the logarithm chosen at quasienergies within the 
$\ell$-th gap.
Ref. \cite{rudner_anomalous_2012} showed the Chern numbers can be inferred from the winding numbers 
by $c_\ell=w_{\ell+1}-w_{\ell}$.

\begin{figure}
\includegraphics[width=2.7in]{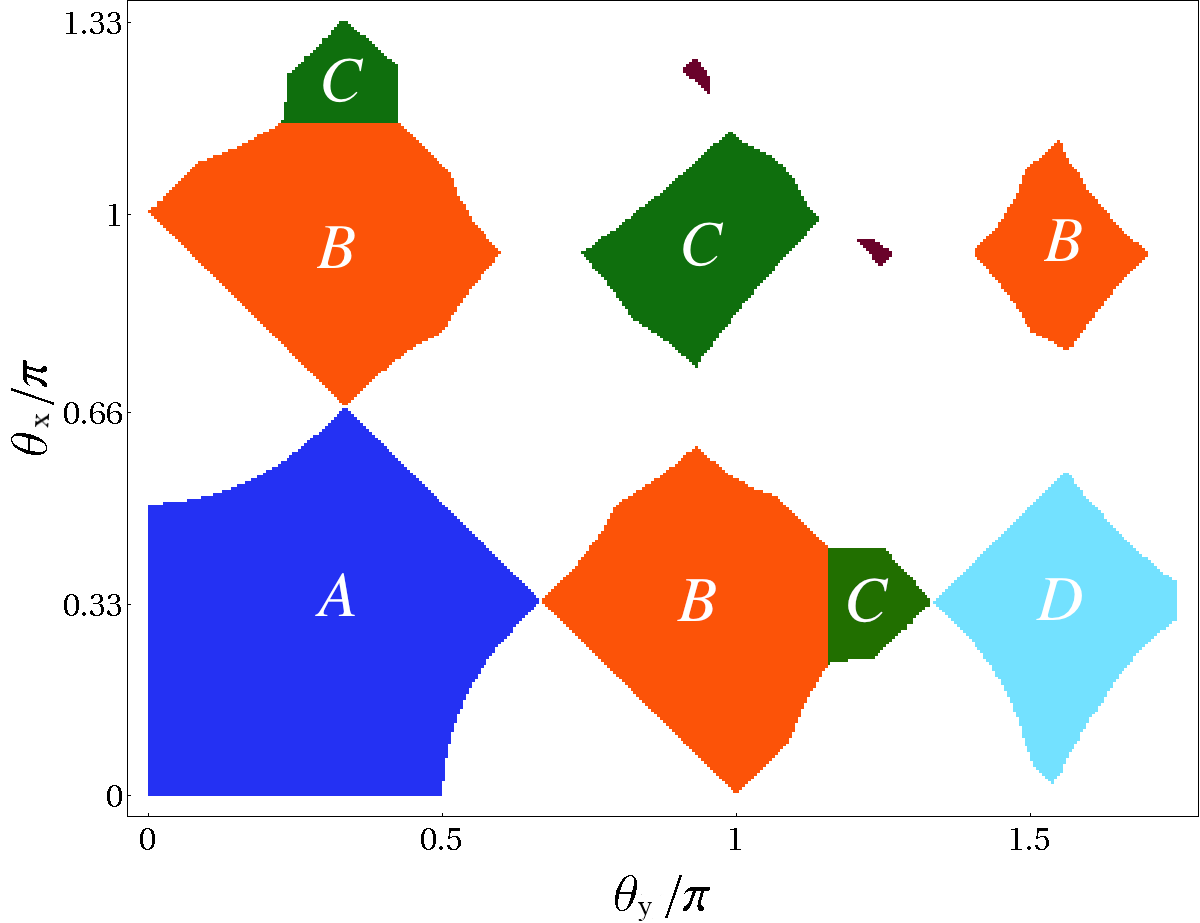}
\includegraphics[width=2.7in]{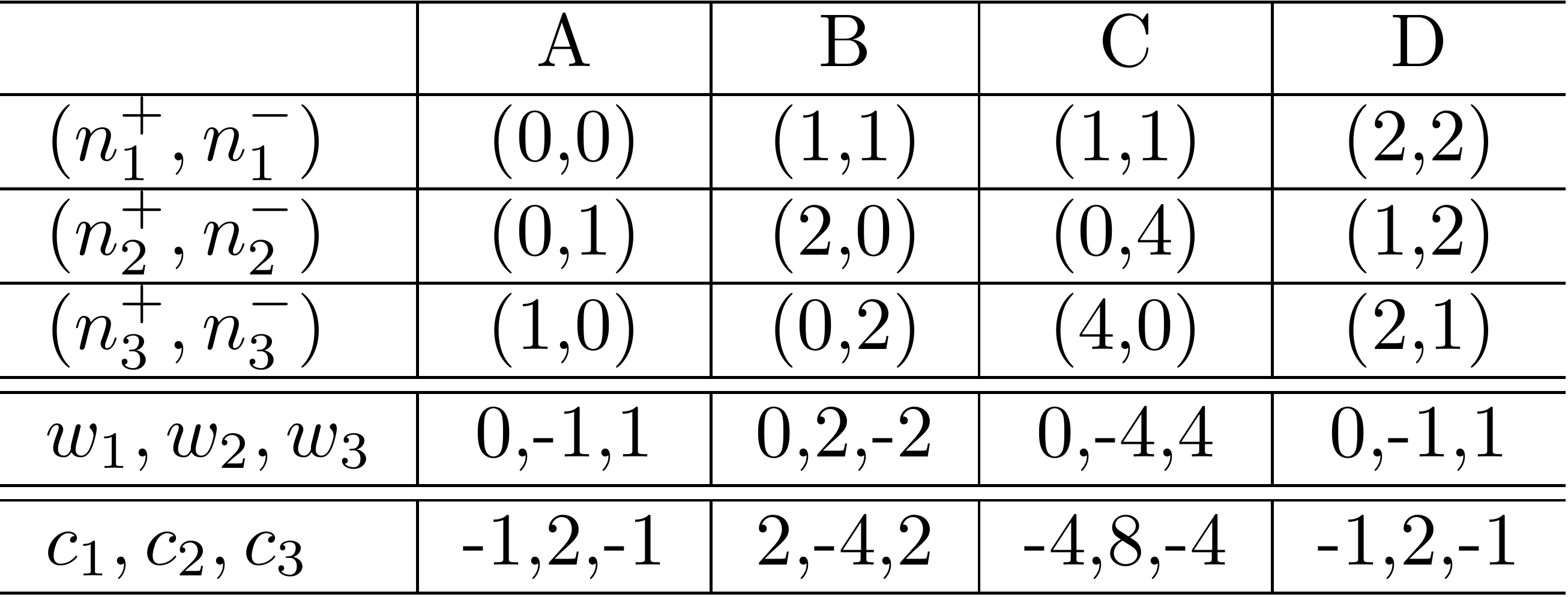}
\caption{(color online) Phase diagram of a periodically driven quantum Hall system in the plane
spanned by driving parameter $\theta_x$ and $\theta_y$ at flux $\alpha=-1/3$. 
Each phase (A, B, C, and D) 
is characterized by $\{(n_{\ell}^+, n_{\ell}^-)\}$, the number of modes within the 
$\ell$-th gap and propagating along 
 $\pm \hat{y}$ at the left edge. The winding number of the $\ell$-th gap $w_\ell = n_{\ell}^+-n_{\ell}^-$, and the Chern number
of the $\ell$-th band
 $c_\ell=w_{\ell+1}-w_{\ell}$ (see main text).
}\label{chern}
\end{figure}

The quasienergy spectra (Fig. 1) manifest a few nice symmetries
of the Floquet operator $U(\mathbf{k},T)$. Related symmetries have been discussed 
for the stationary 
Hofstadter Hamiltonian $H$ \cite{wen1989winding}. 
Firstly, magnetic translational symmetry of $H$ (and $U$)
dictates that an isolated band has $q$-fold
degeneracy for flux $\alpha=p/q$ and its Chern number 
satisfies the Diophantine equation, $pc_\ell+qt_\ell = 1$ where
$t_\ell$ is an integer \cite{dana}. For $p=-1$ and $q=3$, 
$c_\ell=-4,-1,2,5,8$ etc. This forces all the quasienergy
bands to have nonzero Chern numbers differing by multiples of 3.
Secondly, 
$U$ is invariant under spatial inversion $k_y\rightarrow -k_y$, $x\rightarrow L-x$ 
(in the slab geometry). Thus, an edge state solution $\omega(k_y)$ implies another 
edge state at $-k_y$ with the same quasienergy $\omega$ and localized at the opposite edge.
Thirdly, $H$ has a discrete chiral (sublattice) symmetry \cite{PhysRevB.78.195125}: 
$\Gamma H \Gamma=-H$, where $\Gamma$
stands for staggered gauge transformation, 
$\Gamma: | \mathbf{r}\rangle\rightarrow (-1)^{m+n} | \mathbf{r}\rangle$, with $\Gamma^2=1$.
In recipocal space, $\Gamma$ amounts to a $\pi$ shift in $\mathbf{k}$,
$\Gamma H(k_x,k_y) \Gamma=H(k_x+\pi,k_y+\pi)=-H(k_x,k_y)$ \cite{wen1989winding}. 
It follows that for $U$ in the slab geometry,
$
U^*(k_y)=\Gamma_x U(k_y+\pi)\Gamma_x$, 
where operator $\Gamma_x$ performs the local
gauge transformation $|x\rangle \rightarrow (-1)^{x}|x\rangle$.
Therefore, if $\omega(k_y)$ is a quasienergy eigenvalue, 
e.g. an edge state solution,
so is $-\omega$ at shifted momentum $k_y+\pi$.
Two such edge states at $\omega(k_y)$ and $-\omega(k_y+\pi)$ 
reside at the same edge. This will have a significant 
consequence for edge modes at the QBZ boundary,
where $\omega=\pi/T$ and $-\omega=-\pi/T$ 
become equivalent to each other.

Applying the theoretical analysis outlined above, we obtain Fig. 2, the ``phase diagram" 
of the kicked quantum Hall system in terms of two independent
driving parameters, $\theta_x$ and $\theta_y$. 
It showcases four representative phases \footnote{The term phase used here is not to be confused with the many body ground state or the thermodynamic
phase. It refers to parameter regimes of periodically driven systems with characteristic spectral and topological properties.}, labelled by A to D, for flux $\alpha=-1/3$.
All of them feature three well defined quasienergy bands and three gaps, while 
the spectrum in the rest of the phase diagram is largely gapless. 
The corresponding spectrum of each phase in the slab geometry can be found 
in Fig. 1. The table in Fig. 2 summarizes what we know about each phase:
the number of edge modes on the left edge propagating in the $\pm \hat{y}$
direction, $(n_{\ell}^+,n_{\ell}^-)$, inside the $\ell$-th gap; 
the winding number $w_\ell$ of the $\ell$-th gap; and the Chern number $c_\ell$ 
of the $\ell$-th band. Note that $w_\ell$ and $c_\ell$ are calculated independently 
from the bulk spectrum.
At the phase transition points where the gap closes, 
the Chern numbers always change by a multiple of $3$, 
consistent with the Diophantine equation \cite{dana}.
In what follows, we discuss in turn each of these phases.

(A). The main features of phase A can be understood by considering the fast 
driving limit, $\theta_1,\theta_2\ll 1$. The effective
Hamiltonian $\mathscr{H}$, 
takes the same form of $H$
in Eq. \eqref{qh}, only with the bare hopping replaced by the effective hopping
$
J_x\rightarrow J_x\tau/T,\;\; J_y\rightarrow J_y(1-\tau/T).
$
The driven system in phase A stroboscopically mimics 
a static QH system with the same flux but renormalized hopping.
In particular, there is no edge state crossing the gap centered round $\pm \pi/T$.

(B). Phase B 
highlights a remarkable consequence of periodic driving:
there are now two chiral edge modes inside the second and third gap.
This is in sharp contrast to phase A, not only in the number of edge modes,
but also in their chirality. Thus, simple periodic modulations of hopping
proposed here is sufficient to change both the number and the chirality of edge states,
and the Chern numbers of the bands. 
%
%
More importantly, phase B contains a pair of counter-propagating edge modes, dubbed ``$\pi$-modes", inside the first gap 
at the QBZ boundary $\pm\pi/T$. 
These two edge modes, shown in blue for the left edge, have to come in pairs due to the {chiral (sublattice) symmetry} defined above: an edge mode crossing 
the QBZ boundary at $k^a_y$ implies another edge mode also crossing the QBZ boundary 
at $k^b_y=k^a_y+\pi$.
They are guaranteed to have opposite group velocity
because they are related by $\omega(k_y)\leftrightarrow -\omega(k_y+\pi)$.
Such pairs of $\pi$-modes 
are reminiscent of, and of course fundamentally different from, the counter-propagating edge
modes protected by {time-reversal symmetry} in quantum spin Hall effect \cite{PhysRevLett.95.226801}.
The dispersion of the two $\pi$-modes around quasienergy $\pi/T$, labelled by $|\psi_a\rangle=| \uparrow\rangle$ and 
 $|\psi_b\rangle=|\downarrow\rangle$, can be formally described by a 1D Dirac Hamiltonian with chiral symmetry,
$\mathcal{H}_{\pi}=\pi/T-i\sigma_z v_F \partial_y$. Note that $|\psi_{a,b}\rangle=\Gamma|\psi_{b,a}\rangle$, so
 $\Gamma=\sigma_x$ in this basis. After a rotation to a basis where $\Gamma=\sigma_z$
is diagonal, $\mathcal{H}_{\pi}=\pi/T-i\sigma_x v_F \partial_y$, demonstrating that
$\mathcal{H}_{\pi}$ belongs to class AIII$_{(1)}$ of symmetry protected gapless 1D Dirac Hamiltonians 
as classified systematically by Bernard et al \cite{PhysRevB.86.205116}. Thus, perturbations obeying the chiral symmetry,
e.g. small variations in the hopping or the magnetic flux, cannot open a gap \cite{PhysRevB.86.205116}. 

We have further examined the robustness of the $\pi$-modes against static on-site perturbations of the  form
$H'=\sum_{\mathbf{r}}V(\mathbf{r})|\mathbf{r}\rangle\langle\mathbf{r}|$, which break the chiral symmetry.
Kinematically any potential $V$ with a finite Fourier component $V(k_y=\pi)$ tends to mix the two modes. However, 
we find static perturbations including single impurity, staggered potential along $y$, and random disorder potential 
$V(\mathbf{r})\in(-\Delta,\Delta)$ do not open a gap around quasienergy $\pi$. This is verified by
numerically solving for the spectra of finite lattices of dimension $L_x\times L_y$.
To resolve the number of edge states within the first gap,
we define spectral function 
$\rho(k_y,\omega)=\sum_{n,x<L_x/2} \delta(\omega-E_n) |\sum_y\psi_n(x,y) e^{-i k_y y}/{L_y}|^2$,
where the sum over $x$ is restricted to the left half of the slab, 
$E_n$ and $\psi_n$ are the $n$-th quasienergy and the corresponding
eigenwavefunction, respectively. As shown in Fig. 3, $\rho(k_y, \omega)$ 
for $\Delta=0.3J_x$ is peaked at two different $k_y$ values, with a
separation by $\pi$, suggesting two edge modes at and near $\pi/T$ despite the disorder.
These results seem to indicate that the stability of the $\pi$-modes has a topological origin. 
A full understanding however is still lacking. 

Previous work on driven 2D lattice models \cite{kitagawa_topological_2010,rudner_anomalous_2012}
also found chiral edge modes at $\pm\pi/T$. But those $\pi$-modes all have the same chirality.
As a result, the number of  $\pi$-modes can be predicted from the winding number $w_1$,
demonstrating the bulk-boundary correspondence \cite{rudner_anomalous_2012}.
In contrast, here the $\pi$-modes always come in pairs, so the net chirality is zero, $w_1=n^+_1-n^-_1=0$.
The knowledge of the winding or Chern numbers therefore is insufficient to predict the number
or the chirality of the $\pi$-modes.

\begin{figure}
\includegraphics[width=1.8in]{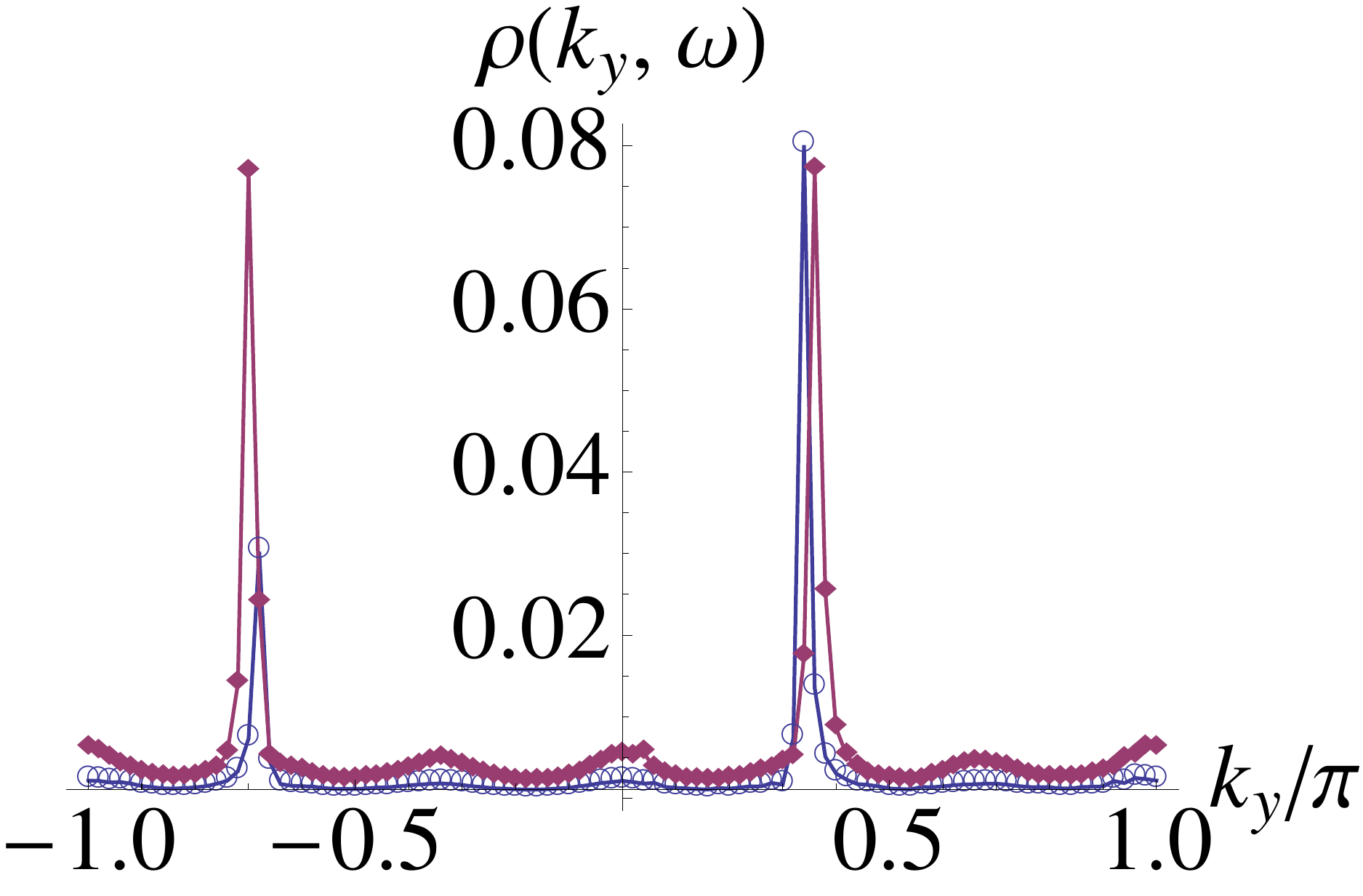}\includegraphics[width=1.4in]{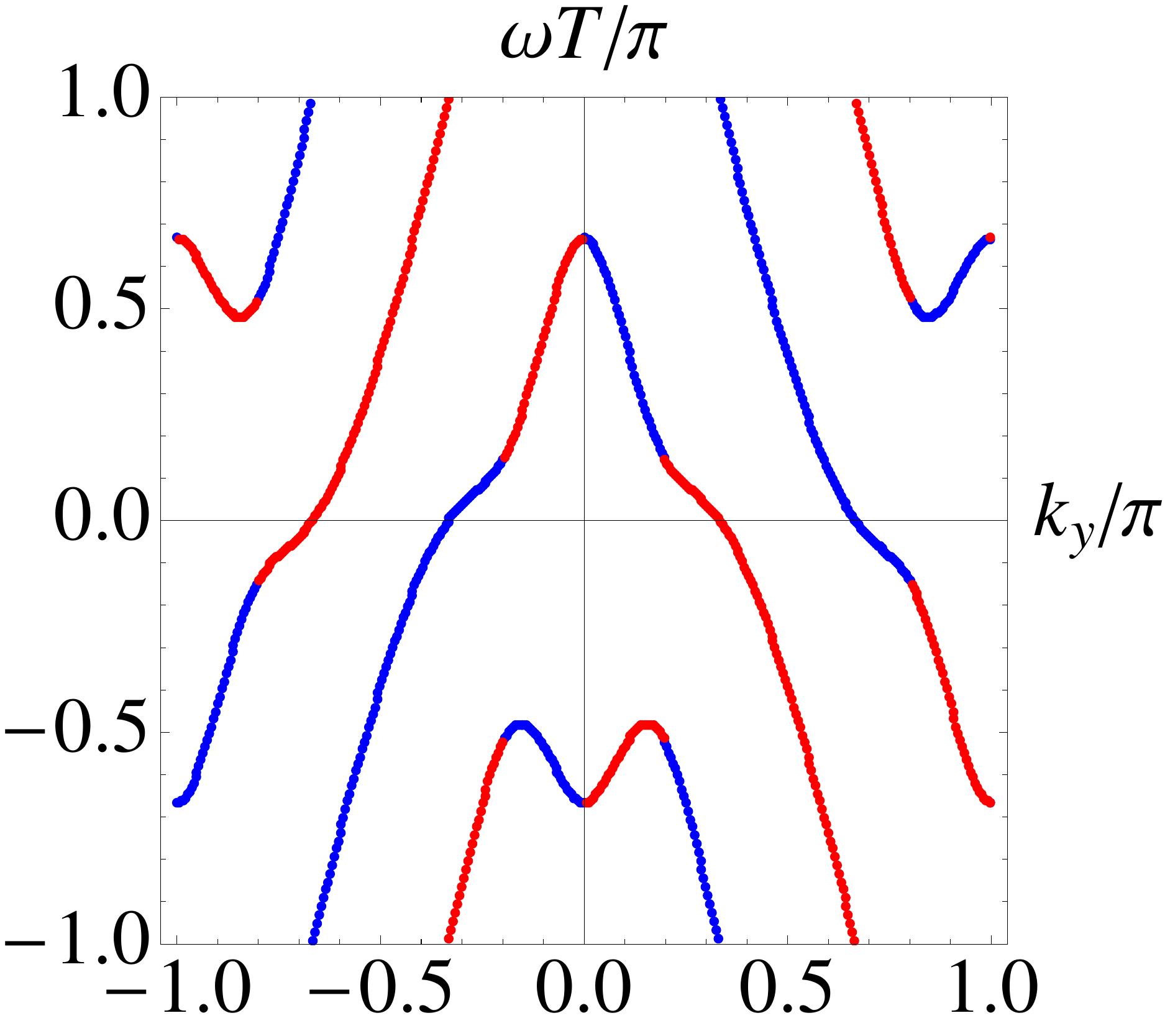}
\caption{(color online) Left: Robust edge states in the presence of disorder 
$\Delta=0.3J_x$. 
The two peaks in the spectral function for $\omega=\pi/T$ (circle)
and $0.9\pi/T$ (filled diamond) suggest 
 two $\pi$-modes at the left edge, consistent with Fig. 1B. 
Right: Winding of the quasienergy spectrum of a 
two-leg ladder.
Red (blue) indicates the eigenstate is predominantly on the right (left) leg.
$\alpha=-1/3$, $\theta_x=\pi/3$, $\theta_y=\pi$.
}\label{finite}
\end{figure}

(C) Phase C is very similar to phase B. 
The only difference is that there 
are 4 (instead of 2 in phase B) chiral edge modes propagating in the same direction 
inside the second and third gap. 
This is yet
another example that Chern numbers of the quasi-energy bands can be 
controlled by periodic driving.

(D) Phase D is qualitatively different from all other phases.
%
Firstly, near the QBZ boundary, there are two pairs of counter-propagating $\pi$-modes,
$n^+_1=n^-_1=2$.
Secondly, the edge states within the second and third gap
also contain counter-propagating modes: two of the edge modes 
propagate in the same direction, but the remaining one propagates in the opposite direction.
For example, $n^+_2=1$, $n^-_2=2$. Although phase D has exactly the same 
set of $\{w_\ell\}$ and $\{c_\ell\}$ as phase A, it has counter-propagating edge
modes in all three quasienergy gaps that are robust against weak disorder. 

The evolution of the edge states and the successive phase transitions 
 as $\theta_{x,y}$ are varied can be captured by a simple model,
a two-leg ladder extending in the $y$ direction.
 For flux $1/3$, the Floquet operator 
of the ladder is $
U(k_y,T)=e^{i\theta_y[-\cos k_y+\sigma_z\sqrt{3}\sin k_y]}e^{i\theta_x\sigma_x}$,
where the $\sigma$'s are the Pauli matrices in the orbital space. 
It follows that the effective Hamiltonian of the ladder
\[
\mathscr{H}(k_y)T=\theta_y\cos k_y+\boldsymbol{\sigma}\cdot\mathbf{h}(k_y),
\]
with 
$h(k_y)=|\mathbf{h}|=$arccos$[\cos \theta_x\cos(\theta_y\sqrt{3}\sin k_y)]$.
Thus, the quasienergy spectrum has two bands (branches), 
\[
\omega_\pm (k_y)T=\theta_y\cos k_y\pm h(k_y),\;\; (\mathrm{mod}\;\; 2\pi).
\]
Figure 3 shows the ladder spectrum for 
$\theta_x=\pi/3$ and $\theta_y=\pi$ (phase B), which agrees remarkably with the edge modes shown in Fig. 1.
As  $\theta_y$ is increased, 
both the curvature and the width of the bands increase. 
Beyond a critical value $\theta_y\simeq 0.57\pi$,
the top of the $\omega_+$ band (and the bottom of the $\omega_-$ band) grows beyond the QBZ, and
re-enters from the opposite side of the QBZ. 
Consequently, the number of states
crossing the QBZ boundary jumps from 0 to 4, marking a transition from phase A to phase B.
From this perspective, the pair of $\pi$-modes results directly  from the
winding of quasienergy  across the QBZ boundary as driving in the $y$-direction 
($\theta_y$) is increased. 
For $\theta_y>1.33\pi$, both the top and bottom of $\omega_\pm$
exceed the QBZ,  giving rise to two pairs of $\pi$-modes at each edge in phase D. When folded into the 
QBZ, they intrude into the second and third gap, leading to
the anomalous edge mode propagating in the ``wrong" direction.

The anomalous edge modes unique to periodically driven QH system can be detected experimentally
by momentum-resolved radio-frequency spectroscopy \cite{Stewart:2008uq}, which measures the spectral function
$\rho(k_y,\omega)$. 
Atoms occupying the $\pi$-mode at quasienergy $\omega$ absorb radio-frequency photon 
and undergo a vertical transition to an empty hyperfine state which can be subsequently imaged. For example, in
phase B, the measured spectral function will feature peaks at $k^{a,b}_y$ and energy $E_n=(2n+1)\pi/T$.
Alternatively, the edge currents can be probed by quantum quenches that convert them into 
density patterns \cite{killi_use_2012} or following the recent proposal of Ref. \cite{Goldman23042013}.


\begin{acknowledgments}
We thank Brandon Anderson, Michael Levin, Xiaopeng Li, Takuya Kitagawa, and Mark Rudner for helpful discussions.
This work is supported by AFOSR (FA9550-12-1-0079), NSF (PHY-1205504), and  NIST (60NANB12D244). EZ also acknowledges the support by NSF PHY11-25915 through KITP. 
\end{acknowledgments}

\bibliography{floquet}

\begin{thebibliography}{29}%
\makeatletter
\providecommand \@ifxundefined [1]{%
 \@ifx{#1\undefined}
}%
\providecommand \@ifnum [1]{%
 \ifnum #1\expandafter \@firstoftwo
 \else \expandafter \@secondoftwo
 \fi
}%
\providecommand \@ifx [1]{%
 \ifx #1\expandafter \@firstoftwo
 \else \expandafter \@secondoftwo
 \fi
}%
\providecommand \natexlab [1]{#1}%
\providecommand \enquote  [1]{``#1''}%
\providecommand \bibnamefont  [1]{#1}%
\providecommand \bibfnamefont [1]{#1}%
\providecommand \citenamefont [1]{#1}%
\providecommand \href@noop [0]{\@secondoftwo}%
\providecommand \href [0]{\begingroup \@sanitize@url \@href}%
\providecommand \@href[1]{\@@startlink{#1}\@@href}%
\providecommand \@@href[1]{\endgroup#1\@@endlink}%
\providecommand \@sanitize@url [0]{\catcode `\\12\catcode `\$12\catcode
  `\&12\catcode `\#12\catcode `\^12\catcode `\_12\catcode `\%12\relax}%
\providecommand \@@startlink[1]{}%
\providecommand \@@endlink[0]{}%
\providecommand \url  [0]{\begingroup\@sanitize@url \@url }%
\providecommand \@url [1]{\endgroup\@href {#1}{\urlprefix }}%
\providecommand \urlprefix  [0]{URL }%
\providecommand \Eprint [0]{\href }%
\providecommand \doibase [0]{http://dx.doi.org/}%
\providecommand \selectlanguage [0]{\@gobble}%
\providecommand \bibinfo  [0]{\@secondoftwo}%
\providecommand \bibfield  [0]{\@secondoftwo}%
\providecommand \translation [1]{[#1]}%
\providecommand \BibitemOpen [0]{}%
\providecommand \bibitemStop [0]{}%
\providecommand \bibitemNoStop [0]{.\EOS\space}%
\providecommand \EOS [0]{\spacefactor3000\relax}%
\providecommand \BibitemShut  [1]{\csname bibitem#1\endcsname}%
\let\auto@bib@innerbib\@empty
\bibitem [{\citenamefont {{Berry}}(1984)}]{1984RSPSA.392.45B}%
  \BibitemOpen
  \bibfield  {author} {\bibinfo {author} {\bibfnamefont {M.~V.}\ \bibnamefont
  {{Berry}}},\ }\href {\doibase 10.1098/rspa.1984.0023} {\bibfield  {journal}
  {\bibinfo  {journal} {Proc. R. Soc. Lond. A}\ }\textbf {\bibinfo {volume}
  {392}},\ \bibinfo {pages} {45} (\bibinfo {year} {1984})}\BibitemShut
  {NoStop}%
\bibitem [{\citenamefont {Aharonov}\ and\ \citenamefont
  {Anandan}(1987)}]{PhysRevLett.58.1593}%
  \BibitemOpen
  \bibfield  {author} {\bibinfo {author} {\bibfnamefont {Y.}~\bibnamefont
  {Aharonov}}\ and\ \bibinfo {author} {\bibfnamefont {J.}~\bibnamefont
  {Anandan}},\ }\href {\doibase 10.1103/PhysRevLett.58.1593} {\bibfield
  {journal} {\bibinfo  {journal} {Phys. Rev. Lett.}\ }\textbf {\bibinfo
  {volume} {58}},\ \bibinfo {pages} {1593} (\bibinfo {year}
  {1987})}\BibitemShut {NoStop}%
\bibitem [{\citenamefont {Oka}\ and\ \citenamefont
  {Aoki}(2009)}]{PhysRevB.79.081406}%
  \BibitemOpen
  \bibfield  {author} {\bibinfo {author} {\bibfnamefont {T.}~\bibnamefont
  {Oka}}\ and\ \bibinfo {author} {\bibfnamefont {H.}~\bibnamefont {Aoki}},\
  }\href {\doibase 10.1103/PhysRevB.79.081406} {\bibfield  {journal} {\bibinfo
  {journal} {Phys. Rev. B}\ }\textbf {\bibinfo {volume} {79}},\ \bibinfo
  {pages} {081406} (\bibinfo {year} {2009})}\BibitemShut {NoStop}%
\bibitem [{\citenamefont {Lindner}\ \emph {et~al.}(2011)\citenamefont
  {Lindner}, \citenamefont {Refael},\ and\ \citenamefont
  {Galitski}}]{lindner_floquet_2011}%
  \BibitemOpen
  \bibfield  {author} {\bibinfo {author} {\bibfnamefont {N.~H.}\ \bibnamefont
  {Lindner}}, \bibinfo {author} {\bibfnamefont {G.}~\bibnamefont {Refael}}, \
  and\ \bibinfo {author} {\bibfnamefont {V.}~\bibnamefont {Galitski}},\ }\href
  {\doibase 10.1038/nphys1926} {\bibfield  {journal} {\bibinfo  {journal}
  {Nature Physics}\ }\textbf {\bibinfo {volume} {7}},\ \bibinfo {pages} {490}
  (\bibinfo {year} {2011})}\BibitemShut {NoStop}%
\bibitem [{\citenamefont {Kitagawa}\ \emph {et~al.}(2010)\citenamefont
  {Kitagawa}, \citenamefont {Berg}, \citenamefont {Rudner},\ and\ \citenamefont
  {Demler}}]{kitagawa_topological_2010}%
  \BibitemOpen
  \bibfield  {author} {\bibinfo {author} {\bibfnamefont {T.}~\bibnamefont
  {Kitagawa}}, \bibinfo {author} {\bibfnamefont {E.}~\bibnamefont {Berg}},
  \bibinfo {author} {\bibfnamefont {M.}~\bibnamefont {Rudner}}, \ and\ \bibinfo
  {author} {\bibfnamefont {E.}~\bibnamefont {Demler}},\ }\href {\doibase
  10.1103/PhysRevB.82.235114} {\bibfield  {journal} {\bibinfo  {journal} {Phys.
  Rev. B}\ }\textbf {\bibinfo {volume} {82}},\ \bibinfo {pages} {235114}
  (\bibinfo {year} {2010})}\BibitemShut {NoStop}%
\bibitem [{\citenamefont {Lindner}\ \emph {et~al.}(2013)\citenamefont
  {Lindner}, \citenamefont {Bergman}, \citenamefont {Refael},\ and\
  \citenamefont {Galitski}}]{lindner_topological_2011}%
  \BibitemOpen
  \bibfield  {author} {\bibinfo {author} {\bibfnamefont {N.~H.}\ \bibnamefont
  {Lindner}}, \bibinfo {author} {\bibfnamefont {D.~L.}\ \bibnamefont
  {Bergman}}, \bibinfo {author} {\bibfnamefont {G.}~\bibnamefont {Refael}}, \
  and\ \bibinfo {author} {\bibfnamefont {V.}~\bibnamefont {Galitski}},\ }\href
  {\doibase 10.1103/PhysRevB.87.235131} {\bibfield  {journal} {\bibinfo
  {journal} {Phys. Rev. B}\ }\textbf {\bibinfo {volume} {87}},\ \bibinfo
  {pages} {235131} (\bibinfo {year} {2013})}\BibitemShut {NoStop}%
\bibitem [{\citenamefont {Kitagawa}\ \emph {et~al.}(2011)\citenamefont
  {Kitagawa}, \citenamefont {Oka}, \citenamefont {Brataas}, \citenamefont
  {Fu},\ and\ \citenamefont {Demler}}]{kitagawa_transport_2011}%
  \BibitemOpen
  \bibfield  {author} {\bibinfo {author} {\bibfnamefont {T.}~\bibnamefont
  {Kitagawa}}, \bibinfo {author} {\bibfnamefont {T.}~\bibnamefont {Oka}},
  \bibinfo {author} {\bibfnamefont {A.}~\bibnamefont {Brataas}}, \bibinfo
  {author} {\bibfnamefont {L.}~\bibnamefont {Fu}}, \ and\ \bibinfo {author}
  {\bibfnamefont {E.}~\bibnamefont {Demler}},\ }\href {\doibase
  10.1103/PhysRevB.84.235108} {\bibfield  {journal} {\bibinfo  {journal} {Phys.
  Rev. B}\ }\textbf {\bibinfo {volume} {84}},\ \bibinfo {pages} {235108}
  (\bibinfo {year} {2011})}\BibitemShut {NoStop}%
\bibitem [{\citenamefont {Jiang}\ \emph {et~al.}(2011)\citenamefont {Jiang},
  \citenamefont {Kitagawa}, \citenamefont {Alicea}, \citenamefont {Akhmerov},
  \citenamefont {Pekker}, \citenamefont {Refael}, \citenamefont {Cirac},
  \citenamefont {Demler}, \citenamefont {Lukin},\ and\ \citenamefont
  {Zoller}}]{jiang_majorana_2011}%
  \BibitemOpen
  \bibfield  {author} {\bibinfo {author} {\bibfnamefont {L.}~\bibnamefont
  {Jiang}}, \bibinfo {author} {\bibfnamefont {T.}~\bibnamefont {Kitagawa}},
  \bibinfo {author} {\bibfnamefont {J.}~\bibnamefont {Alicea}}, \bibinfo
  {author} {\bibfnamefont {A.~R.}\ \bibnamefont {Akhmerov}}, \bibinfo {author}
  {\bibfnamefont {D.}~\bibnamefont {Pekker}}, \bibinfo {author} {\bibfnamefont
  {G.}~\bibnamefont {Refael}}, \bibinfo {author} {\bibfnamefont {J.~I.}\
  \bibnamefont {Cirac}}, \bibinfo {author} {\bibfnamefont {E.}~\bibnamefont
  {Demler}}, \bibinfo {author} {\bibfnamefont {M.~D.}\ \bibnamefont {Lukin}}, \
  and\ \bibinfo {author} {\bibfnamefont {P.}~\bibnamefont {Zoller}},\ }\href
  {\doibase 10.1103/PhysRevLett.106.220402} {\bibfield  {journal} {\bibinfo
  {journal} {Phys. Rev. Lett.}\ }\textbf {\bibinfo {volume} {106}},\ \bibinfo
  {pages} {220402} (\bibinfo {year} {2011})}\BibitemShut {NoStop}%
\bibitem [{\citenamefont {Gu}\ \emph {et~al.}(2011)\citenamefont {Gu},
  \citenamefont {Fertig}, \citenamefont {Arovas},\ and\ \citenamefont
  {Auerbach}}]{gu_floquet_2011}%
  \BibitemOpen
  \bibfield  {author} {\bibinfo {author} {\bibfnamefont {Z.}~\bibnamefont
  {Gu}}, \bibinfo {author} {\bibfnamefont {H.~A.}\ \bibnamefont {Fertig}},
  \bibinfo {author} {\bibfnamefont {D.~P.}\ \bibnamefont {Arovas}}, \ and\
  \bibinfo {author} {\bibfnamefont {A.}~\bibnamefont {Auerbach}},\ }\href
  {\doibase 10.1103/PhysRevLett.107.216601} {\bibfield  {journal} {\bibinfo
  {journal} {Phys. Rev. Lett.}\ }\textbf {\bibinfo {volume} {107}},\ \bibinfo
  {pages} {216601} (\bibinfo {year} {2011})}\BibitemShut {NoStop}%
\bibitem [{\citenamefont {Rudner}\ \emph {et~al.}(2013)\citenamefont {Rudner},
  \citenamefont {Lindner}, \citenamefont {Berg},\ and\ \citenamefont
  {Levin}}]{rudner_anomalous_2012}%
  \BibitemOpen
  \bibfield  {author} {\bibinfo {author} {\bibfnamefont {M.~S.}\ \bibnamefont
  {Rudner}}, \bibinfo {author} {\bibfnamefont {N.~H.}\ \bibnamefont {Lindner}},
  \bibinfo {author} {\bibfnamefont {E.}~\bibnamefont {Berg}}, \ and\ \bibinfo
  {author} {\bibfnamefont {M.}~\bibnamefont {Levin}},\ }\href {\doibase
  10.1103/PhysRevX.3.031005} {\bibfield  {journal} {\bibinfo  {journal} {Phys.
  Rev. X}\ }\textbf {\bibinfo {volume} {3}},\ \bibinfo {pages} {031005}
  (\bibinfo {year} {2013})}\BibitemShut {NoStop}%
\bibitem [{\citenamefont {{Thakurathi}}\ \emph {et~al.}(2013)\citenamefont
  {{Thakurathi}}, \citenamefont {{Patel}}, \citenamefont {{Sen}},\ and\
  \citenamefont {{Dutta}}}]{2013arXiv1303.2300T}%
  \BibitemOpen
  \bibfield  {author} {\bibinfo {author} {\bibfnamefont {M.}~\bibnamefont
  {{Thakurathi}}}, \bibinfo {author} {\bibfnamefont {A.~A.}\ \bibnamefont
  {{Patel}}}, \bibinfo {author} {\bibfnamefont {D.}~\bibnamefont {{Sen}}}, \
  and\ \bibinfo {author} {\bibfnamefont {A.}~\bibnamefont {{Dutta}}},\
  }\href@noop {} {} (\bibinfo {year} {2013}),\ \Eprint
  {http://arxiv.org/abs/1303.2300} {arXiv:1303.2300} \BibitemShut {NoStop}%
\bibitem [{\citenamefont {Rechtsman}\ \emph {et~al.}(2013)\citenamefont
  {Rechtsman}, \citenamefont {Zeuner}, \citenamefont {Plotnik}, \citenamefont
  {Lumer}, \citenamefont {Podolsky}, \citenamefont {Dreisow}, \citenamefont
  {Nolte}, \citenamefont {Segev},\ and\ \citenamefont
  {Szameit}}]{Rechtsman:2013fk}%
  \BibitemOpen
  \bibfield  {author} {\bibinfo {author} {\bibfnamefont {M.~C.}\ \bibnamefont
  {Rechtsman}}, \bibinfo {author} {\bibfnamefont {J.~M.}\ \bibnamefont
  {Zeuner}}, \bibinfo {author} {\bibfnamefont {Y.}~\bibnamefont {Plotnik}},
  \bibinfo {author} {\bibfnamefont {Y.}~\bibnamefont {Lumer}}, \bibinfo
  {author} {\bibfnamefont {D.}~\bibnamefont {Podolsky}}, \bibinfo {author}
  {\bibfnamefont {F.}~\bibnamefont {Dreisow}}, \bibinfo {author} {\bibfnamefont
  {S.}~\bibnamefont {Nolte}}, \bibinfo {author} {\bibfnamefont
  {M.}~\bibnamefont {Segev}}, \ and\ \bibinfo {author} {\bibfnamefont
  {A.}~\bibnamefont {Szameit}},\ }\href {http://dx.doi.org/10.1038/nature12066}
  {\bibfield  {journal} {\bibinfo  {journal} {Nature}\ }\textbf {\bibinfo
  {volume} {496}},\ \bibinfo {pages} {196} (\bibinfo {year}
  {2013})}\BibitemShut {NoStop}%
\bibitem [{\citenamefont {Thouless}(1983)}]{PhysRevB.27.6083}%
  \BibitemOpen
  \bibfield  {author} {\bibinfo {author} {\bibfnamefont {D.~J.}\ \bibnamefont
  {Thouless}},\ }\href {\doibase 10.1103/PhysRevB.27.6083} {\bibfield
  {journal} {\bibinfo  {journal} {Phys. Rev. B}\ }\textbf {\bibinfo {volume}
  {27}},\ \bibinfo {pages} {6083} (\bibinfo {year} {1983})}\BibitemShut
  {NoStop}%
\bibitem [{\citenamefont {Kitaev}(2009)}]{kitaev}%
  \BibitemOpen
  \bibfield  {author} {\bibinfo {author} {\bibfnamefont {A.}~\bibnamefont
  {Kitaev}},\ }\href@noop {} {\bibfield  {journal} {\bibinfo  {journal} {AIP
  Conf. Proc.}\ }\textbf {\bibinfo {volume} {1134}},\ \bibinfo {pages} {22}
  (\bibinfo {year} {2009})}\BibitemShut {NoStop}%
\bibitem [{\citenamefont {Schnyder}\ \emph {et~al.}(2008)\citenamefont
  {Schnyder}, \citenamefont {Ryu}, \citenamefont {Furusaki},\ and\
  \citenamefont {Ludwig}}]{PhysRevB.78.195125}%
  \BibitemOpen
  \bibfield  {author} {\bibinfo {author} {\bibfnamefont {A.~P.}\ \bibnamefont
  {Schnyder}}, \bibinfo {author} {\bibfnamefont {S.}~\bibnamefont {Ryu}},
  \bibinfo {author} {\bibfnamefont {A.}~\bibnamefont {Furusaki}}, \ and\
  \bibinfo {author} {\bibfnamefont {A.~W.~W.}\ \bibnamefont {Ludwig}},\ }\href
  {\doibase 10.1103/PhysRevB.78.195125} {\bibfield  {journal} {\bibinfo
  {journal} {Phys. Rev. B}\ }\textbf {\bibinfo {volume} {78}},\ \bibinfo
  {pages} {195125} (\bibinfo {year} {2008})}\BibitemShut {NoStop}%
\bibitem [{\citenamefont {Lin}\ \emph {et~al.}(2009)\citenamefont {Lin},
  \citenamefont {Compton}, \citenamefont {Jimenez-Garcia}, \citenamefont
  {Porto},\ and\ \citenamefont {Spielman}}]{Lin:2009fk}%
  \BibitemOpen
  \bibfield  {author} {\bibinfo {author} {\bibfnamefont {Y.~J.}\ \bibnamefont
  {Lin}}, \bibinfo {author} {\bibfnamefont {R.~L.}\ \bibnamefont {Compton}},
  \bibinfo {author} {\bibfnamefont {K.}~\bibnamefont {Jimenez-Garcia}},
  \bibinfo {author} {\bibfnamefont {J.~V.}\ \bibnamefont {Porto}}, \ and\
  \bibinfo {author} {\bibfnamefont {I.~B.}\ \bibnamefont {Spielman}},\ }\href
  {http://dx.doi.org/10.1038/nature08609} {\bibfield  {journal} {\bibinfo
  {journal} {Nature}\ }\textbf {\bibinfo {volume} {462}},\ \bibinfo {pages}
  {628} (\bibinfo {year} {2009})}\BibitemShut {NoStop}%
\bibitem [{\citenamefont {Aidelsburger}\ \emph {et~al.}(2011)\citenamefont
  {Aidelsburger}, \citenamefont {Atala}, \citenamefont {Nascimb\`ene},
  \citenamefont {Trotzky}, \citenamefont {Chen},\ and\ \citenamefont
  {Bloch}}]{PhysRevLett.107.255301}%
  \BibitemOpen
  \bibfield  {author} {\bibinfo {author} {\bibfnamefont {M.}~\bibnamefont
  {Aidelsburger}}, \bibinfo {author} {\bibfnamefont {M.}~\bibnamefont {Atala}},
  \bibinfo {author} {\bibfnamefont {S.}~\bibnamefont {Nascimb\`ene}}, \bibinfo
  {author} {\bibfnamefont {S.}~\bibnamefont {Trotzky}}, \bibinfo {author}
  {\bibfnamefont {Y.-A.}\ \bibnamefont {Chen}}, \ and\ \bibinfo {author}
  {\bibfnamefont {I.}~\bibnamefont {Bloch}},\ }\href {\doibase
  10.1103/PhysRevLett.107.255301} {\bibfield  {journal} {\bibinfo  {journal}
  {Phys. Rev. Lett.}\ }\textbf {\bibinfo {volume} {107}},\ \bibinfo {pages}
  {255301} (\bibinfo {year} {2011})}\BibitemShut {NoStop}%
\bibitem [{\citenamefont {Struck}\ \emph {et~al.}(2012)\citenamefont {Struck},
  \citenamefont {\"Olschl\"ager}, \citenamefont {Weinberg}, \citenamefont
  {Hauke}, \citenamefont {Simonet}, \citenamefont {Eckardt}, \citenamefont
  {Lewenstein}, \citenamefont {Sengstock},\ and\ \citenamefont
  {Windpassinger}}]{PhysRevLett.108.225304}%
  \BibitemOpen
  \bibfield  {author} {\bibinfo {author} {\bibfnamefont {J.}~\bibnamefont
  {Struck}}, \bibinfo {author} {\bibfnamefont {C.}~\bibnamefont
  {\"Olschl\"ager}}, \bibinfo {author} {\bibfnamefont {M.}~\bibnamefont
  {Weinberg}}, \bibinfo {author} {\bibfnamefont {P.}~\bibnamefont {Hauke}},
  \bibinfo {author} {\bibfnamefont {J.}~\bibnamefont {Simonet}}, \bibinfo
  {author} {\bibfnamefont {A.}~\bibnamefont {Eckardt}}, \bibinfo {author}
  {\bibfnamefont {M.}~\bibnamefont {Lewenstein}}, \bibinfo {author}
  {\bibfnamefont {K.}~\bibnamefont {Sengstock}}, \ and\ \bibinfo {author}
  {\bibfnamefont {P.}~\bibnamefont {Windpassinger}},\ }\href {\doibase
  10.1103/PhysRevLett.108.225304} {\bibfield  {journal} {\bibinfo  {journal}
  {Phys. Rev. Lett.}\ }\textbf {\bibinfo {volume} {108}},\ \bibinfo {pages}
  {225304} (\bibinfo {year} {2012})}\BibitemShut {NoStop}%
\bibitem [{\citenamefont {Hauke}\ \emph {et~al.}(2012)\citenamefont {Hauke},
  \citenamefont {Tieleman}, \citenamefont {Celi}, \citenamefont
  {\"Olschl\"ager}, \citenamefont {Simonet}, \citenamefont {Struck},
  \citenamefont {Weinberg}, \citenamefont {Windpassinger}, \citenamefont
  {Sengstock}, \citenamefont {Lewenstein},\ and\ \citenamefont
  {Eckardt}}]{PhysRevLett.109.145301}%
  \BibitemOpen
  \bibfield  {author} {\bibinfo {author} {\bibfnamefont {P.}~\bibnamefont
  {Hauke}}, \bibinfo {author} {\bibfnamefont {O.}~\bibnamefont {Tieleman}},
  \bibinfo {author} {\bibfnamefont {A.}~\bibnamefont {Celi}}, \bibinfo {author}
  {\bibfnamefont {C.}~\bibnamefont {\"Olschl\"ager}}, \bibinfo {author}
  {\bibfnamefont {J.}~\bibnamefont {Simonet}}, \bibinfo {author} {\bibfnamefont
  {J.}~\bibnamefont {Struck}}, \bibinfo {author} {\bibfnamefont
  {M.}~\bibnamefont {Weinberg}}, \bibinfo {author} {\bibfnamefont
  {P.}~\bibnamefont {Windpassinger}}, \bibinfo {author} {\bibfnamefont
  {K.}~\bibnamefont {Sengstock}}, \bibinfo {author} {\bibfnamefont
  {M.}~\bibnamefont {Lewenstein}}, \ and\ \bibinfo {author} {\bibfnamefont
  {A.}~\bibnamefont {Eckardt}},\ }\href {\doibase
  10.1103/PhysRevLett.109.145301} {\bibfield  {journal} {\bibinfo  {journal}
  {Phys. Rev. Lett.}\ }\textbf {\bibinfo {volume} {109}},\ \bibinfo {pages}
  {145301} (\bibinfo {year} {2012})}\BibitemShut {NoStop}%
\bibitem [{\citenamefont {Hofstadter}(1976)}]{hofstadter_energy_1976}%
  \BibitemOpen
  \bibfield  {author} {\bibinfo {author} {\bibfnamefont {D.~R.}\ \bibnamefont
  {Hofstadter}},\ }\href {\doibase 10.1103/PhysRevB.14.2239} {\bibfield
  {journal} {\bibinfo  {journal} {Phys. Rev. B}\ }\textbf {\bibinfo {volume}
  {14}},\ \bibinfo {pages} {2239} (\bibinfo {year} {1976})}\BibitemShut
  {NoStop}%
\bibitem [{\citenamefont {Thouless}\ \emph {et~al.}(1982)\citenamefont
  {Thouless}, \citenamefont {Kohmoto}, \citenamefont {Nightingale},\ and\
  \citenamefont {den Nijs}}]{thouless_quantized_1982}%
  \BibitemOpen
  \bibfield  {author} {\bibinfo {author} {\bibfnamefont {D.~J.}\ \bibnamefont
  {Thouless}}, \bibinfo {author} {\bibfnamefont {M.}~\bibnamefont {Kohmoto}},
  \bibinfo {author} {\bibfnamefont {M.~P.}\ \bibnamefont {Nightingale}}, \ and\
  \bibinfo {author} {\bibfnamefont {M.}~\bibnamefont {den Nijs}},\ }\href
  {\doibase 10.1103/PhysRevLett.49.405} {\bibfield  {journal} {\bibinfo
  {journal} {Phys. Rev. Lett.}\ }\textbf {\bibinfo {volume} {49}},\ \bibinfo
  {pages} {405} (\bibinfo {year} {1982})}\BibitemShut {NoStop}%
\bibitem [{\citenamefont {Wen}\ and\ \citenamefont
  {Zee}(1989)}]{wen1989winding}%
  \BibitemOpen
  \bibfield  {author} {\bibinfo {author} {\bibfnamefont {X.}~\bibnamefont
  {Wen}}\ and\ \bibinfo {author} {\bibfnamefont {A.}~\bibnamefont {Zee}},\
  }\href@noop {} {\bibfield  {journal} {\bibinfo  {journal} {Nuclear Physics
  B}\ }\textbf {\bibinfo {volume} {316}},\ \bibinfo {pages} {641} (\bibinfo
  {year} {1989})}\BibitemShut {NoStop}%
\bibitem [{\citenamefont {Dana}\ \emph {et~al.}(1985)\citenamefont {Dana},
  \citenamefont {Avron},\ and\ \citenamefont {Zak}}]{dana}%
  \BibitemOpen
  \bibfield  {author} {\bibinfo {author} {\bibfnamefont {I.}~\bibnamefont
  {Dana}}, \bibinfo {author} {\bibfnamefont {Y.}~\bibnamefont {Avron}}, \ and\
  \bibinfo {author} {\bibfnamefont {J.}~\bibnamefont {Zak}},\ }\href
  {http://stacks.iop.org/0022-3719/18/i=22/a=004} {\bibfield  {journal}
  {\bibinfo  {journal} {J. Phys. C}\ }\textbf {\bibinfo {volume} {18}},\
  \bibinfo {pages} {L679} (\bibinfo {year} {1985})}\BibitemShut {NoStop}%
\bibitem [{Note1()}]{Note1}%
  \BibitemOpen
  \bibinfo {note} {The term phase used here is not to be confused with the many
  body ground state or the thermodynamic phase. It refers to parameter regimes
  of periodically driven systems with characteristic spectral and topological
  properties.}\BibitemShut {Stop}%
\bibitem [{\citenamefont {Kane}\ and\ \citenamefont
  {Mele}(2005)}]{PhysRevLett.95.226801}%
  \BibitemOpen
  \bibfield  {author} {\bibinfo {author} {\bibfnamefont {C.~L.}\ \bibnamefont
  {Kane}}\ and\ \bibinfo {author} {\bibfnamefont {E.~J.}\ \bibnamefont
  {Mele}},\ }\href {\doibase 10.1103/PhysRevLett.95.226801} {\bibfield
  {journal} {\bibinfo  {journal} {Phys. Rev. Lett.}\ }\textbf {\bibinfo
  {volume} {95}},\ \bibinfo {pages} {226801} (\bibinfo {year}
  {2005})}\BibitemShut {NoStop}%
\bibitem [{\citenamefont {Bernard}\ \emph {et~al.}(2012)\citenamefont
  {Bernard}, \citenamefont {Kim},\ and\ \citenamefont
  {LeClair}}]{PhysRevB.86.205116}%
  \BibitemOpen
  \bibfield  {author} {\bibinfo {author} {\bibfnamefont {D.}~\bibnamefont
  {Bernard}}, \bibinfo {author} {\bibfnamefont {E.-A.}\ \bibnamefont {Kim}}, \
  and\ \bibinfo {author} {\bibfnamefont {A.}~\bibnamefont {LeClair}},\ }\href
  {\doibase 10.1103/PhysRevB.86.205116} {\bibfield  {journal} {\bibinfo
  {journal} {Phys. Rev. B}\ }\textbf {\bibinfo {volume} {86}},\ \bibinfo
  {pages} {205116} (\bibinfo {year} {2012})}\BibitemShut {NoStop}%
\bibitem [{\citenamefont {Stewart}\ \emph {et~al.}(2008)\citenamefont
  {Stewart}, \citenamefont {Gaebler},\ and\ \citenamefont
  {Jin}}]{Stewart:2008uq}%
  \BibitemOpen
  \bibfield  {author} {\bibinfo {author} {\bibfnamefont {J.~T.}\ \bibnamefont
  {Stewart}}, \bibinfo {author} {\bibfnamefont {J.~P.}\ \bibnamefont
  {Gaebler}}, \ and\ \bibinfo {author} {\bibfnamefont {D.~S.}\ \bibnamefont
  {Jin}},\ }\href {http://dx.doi.org/10.1038/nature07172} {\bibfield  {journal}
  {\bibinfo  {journal} {Nature}\ }\textbf {\bibinfo {volume} {454}},\ \bibinfo
  {pages} {744} (\bibinfo {year} {2008})}\BibitemShut {NoStop}%
\bibitem [{\citenamefont {Killi}\ and\ \citenamefont
  {Paramekanti}(2012)}]{killi_use_2012}%
  \BibitemOpen
  \bibfield  {author} {\bibinfo {author} {\bibfnamefont {M.}~\bibnamefont
  {Killi}}\ and\ \bibinfo {author} {\bibfnamefont {A.}~\bibnamefont
  {Paramekanti}},\ }\href {\doibase 10.1103/PhysRevA.85.061606} {\bibfield
  {journal} {\bibinfo  {journal} {Phys. Rev. A}\ }\textbf {\bibinfo {volume}
  {85}},\ \bibinfo {pages} {061606} (\bibinfo {year} {2012})}\BibitemShut
  {NoStop}%
\bibitem [{\citenamefont {Goldman}\ \emph {et~al.}(2013)\citenamefont
  {Goldman}, \citenamefont {Dalibard}, \citenamefont {Dauphin}, \citenamefont
  {Gerbier}, \citenamefont {Lewenstein}, \citenamefont {Zoller},\ and\
  \citenamefont {Spielman}}]{Goldman23042013}%
  \BibitemOpen
  \bibfield  {author} {\bibinfo {author} {\bibfnamefont {N.}~\bibnamefont
  {Goldman}}, \bibinfo {author} {\bibfnamefont {J.}~\bibnamefont {Dalibard}},
  \bibinfo {author} {\bibfnamefont {A.}~\bibnamefont {Dauphin}}, \bibinfo
  {author} {\bibfnamefont {F.}~\bibnamefont {Gerbier}}, \bibinfo {author}
  {\bibfnamefont {M.}~\bibnamefont {Lewenstein}}, \bibinfo {author}
  {\bibfnamefont {P.}~\bibnamefont {Zoller}}, \ and\ \bibinfo {author}
  {\bibfnamefont {I.~B.}\ \bibnamefont {Spielman}},\ }\href {\doibase
  10.1073/pnas.1300170110} {\bibfield  {journal} {\bibinfo  {journal} {Proc.
  Natl. Acad. Sci.}\ }\textbf {\bibinfo {volume} {110}},\ \bibinfo {pages}
  {6736} (\bibinfo {year} {2013})}\BibitemShut {NoStop}%
\end{thebibliography}%

\end{document}